\documentclass{article}

    \usepackage[final]{neurips_2024}

\usepackage[utf8]{inputenc} 
\usepackage[T1]{fontenc}    
\usepackage{hyperref}       
\usepackage{url}            
\usepackage{booktabs}       
\usepackage{amsfonts}       
\usepackage{nicefrac}       
\usepackage{microtype}      
\usepackage{xcolor}         
\usepackage{algorithm}
\usepackage{algpseudocode}
\usepackage{amsmath}
\usepackage{graphicx}
\usepackage{subfig}
\usepackage{wrapfig}

\title{Exploratory Study Of Human-AI Interaction For Hindustani Music}

\author{%
  Nithya Shikarpur \\
  Mila - Quebec Artificial Intelligence Institute, Université de Montréal\\
  \texttt{snnithya@mit.edu} \\
  \And
  Cheng-Zhi Anna Huang \\
  Mila - Quebec Artificial Intelligence Institute, Université de Montréal, Canada CIFAR AI Chair\\
  \texttt{anna.huang@mila.quebec}
}

\begin{document}

\maketitle

\begin{abstract}
  This paper presents a study of participants interacting with and using GaMaDHaNi, a novel hierarchical generative model for Hindustani vocal contours. To explore possible use cases in human-AI interaction, we conducted a user study with three participants, each engaging with the model through three predefined interaction modes. Although this study was conducted "in the wild"— with the model unadapted for the shift from the training data to real-world interaction — we use it as a pilot to better understand the expectations, reactions, and preferences of practicing musicians when engaging with such a model. We note their challenges as (1) the lack of restrictions in model output, and (2) the incoherence of model output. We situate these challenges in the context of Hindustani music and aim to suggest future directions for the model design to address these gaps.\footnotemark
\end{abstract}
\footnotetext[1]{Supplementary video samples available at \url{https://cedar-decade-974.notion.site/Example-Videos-From-User-Studies-b0a17be4f5ed4f3184cc1d110f450541?pvs=74}}

\section{Introduction}

Hindustani music is a form of North Indian classical music that incorporates improvisation guided by fixed melodic (\textit{raga}), rhythmic (\textit{taal}), and structural frameworks. With a focus on improvisation, the tradition has evolved to be predominantly oral; thus a discrete symbolic representation of the music, although present, does not fully capture the melodic ornamentations or the essence of improvisation present in the music. As a result, musical interaction as opposed to textual knowledge is important for both pedagogy and performance of this tradition.

As an oral tradition, Hindustani music is known to be \textit{Gurumukhi}, i.e. to be learned from the \textit{guru} or teacher (\cite{patel-2007, ganguli2012guruji}). Real-time musical interaction with the \textit{guru} is vital for the student to internalize concepts like \textit{raga} boundaries, aesthetics, and vocabulary, which in turn shape their melodic improvisation. Additionally, in performance, a main artist, rhythmic accompaniment, and sometimes melodic accompaniment collaborate, with the accompanists supporting the main artist’s improvised creations, resulting in complex musical interactions (\cite{leante2016observing}).

Considering the importance of interaction in Hindustani music, an interesting area of study is the feasibility of human-AI interaction in this field. Previous work on generative modeling for this music has side-stepped the lack of well-defined discrete representations with two methods: (1) using musical notation from textbooks or music theory (\cite{airaga, das2005finite, Sahasrabuddhe}), (2) leveraging MIDI extracted from audio (\cite{gopi_introductory_2023, automaticGenAdhikary}). However, both methods ignore the rich melodic ornamentation present in the music. GaMaDHaNi (\cite{gamadhani}), a modular hierarchical generative model for Hindustani vocal melodies addresses this gap by employing a two-level hierarchy of data representation including continuous fundamental frequency contours, further referred to as pitch contours, and spectrograms. The model's ability to generate Hindustani-sounding and controllable musical sequences, achieved by capturing the inherent continuous melodic movements through an interpretable intermediate pitch contour representation, renders it a promising candidate for human-AI interaction studies.

In this paper, we gauge the expectations, difficulties, and preferences of participant musicians when interacting with GaMaDHaNi (\cite{gamadhani}). We present:
\begin{itemize}
    \item A preliminary exploratory user study to gauge the quality and feasibility of interaction with GaMaDHaNi, a generative model for Hindustani vocal contours.
    \item A discussion on the findings from the study and suggestions for future improvements.
\end{itemize}

\section{Study Design}
Below we discuss details of the participants involved and the structure of the user study.
\subsection{Study Participants}
We recruited three musicians for the study, further referred to as P1, P2, and P3 due to anonymity purposes. All participants are trained in Hindustani music for over 15 years and engage in regular practice and performance. 
P1 and P2 are harmonium\footnote{Harmonium is an Indian modification of the European free-reed, keyboard aerophone which is commonly used for melodic accompaniment in Hindustani music.} players and P3 is a vocalist. In the interview, P1 interacted with the model using his harmonium whereas P2 and P3 used their voices. The interview lasted for about an hour and the participants were offered a compensation of 20 CAD for their time. 

\subsection{Study Structure}

This study investigates the expectations and perspectives of a Hindustani music practitioner regarding generative tools like GaMaDHaNi, assessing the model's alignment with these insights and identifying its strengths and limitations. As one of the pioneering works in modeling Hindustani vocal contours, GaMaDHaNi lays the groundwork for future research. Through these interviews, we aim to uncover promising directions for researchers and musicians exploring human-AI interaction in the context of Hindustani music. The study consisted of three stages: (1) a semi-structured interview on the participant's relationship with AI, (2) interaction with the model via three pre-defined tasks, and (3) closing reflections.

\textbf{Semi-structured interview}\space\space The study began by exploring the participant's familiarity with generative AI tools and their ideas for how it could assist or enhance their music practice and performance.

\textbf{Interaction with the model}\space\space In this section, we explore the participants' experience interacting with GaMaDHaNi through three pre-defined tasks, tailored to the model's affordances and the potential for engaging interesting human-AI musical interactions. Participants could engage with each task for any duration and were asked to share their experiences afterward.   

The three pre-defined tasks included:
\begin{itemize}
    \item \textbf{Idea Generation} (Task 1): In this task, participants began by selecting their preferred option from two randomly generated, four-second audio clips produced by the model. They then continued to choose their preferred continuation after each generation, with the model providing two new, two-second continuations for each selected clip in an iterative loop.
    \item \textbf{Call and Response} (Task 2): Participants provided a musical input by singing or playing an instrument. In response, GaMaDHaNi, trained to generate 12-second audio clips, produced an eight-second continuation based on the last four seconds of the input.
    \item \textbf{Melodic Reinterpretation} (Task 3): Similar to SDEdit's (\cite{meng2022sdedit}) method to convert coarse brush strokes to full images, in this task, the user's input was considered to be a `guide' based on which the model generated an audio clip, described further in Sec. \ref{subsec:melodic-reinterp}. 
\end{itemize}

\textbf{Closing Thoughts}\space\space Participants were asked for any final thoughts in an open discussion.

The studies were documented using video and screen recordings after obtaining participants' consent. A thematic analysis (\cite{braun2006using}) was performed on the interview transcripts. This paper primarily discusses participant interactions with the model (stage 2), incorporating insights from the other two stages of the study when relevant.

\section{Overview of the Generative Model (GaMaDHaNi)}

\begin{wrapfigure}{r}{0.6\textwidth}
    \centering 
    \vspace{-10mm}
    \includegraphics[width=0.6\textwidth, trim={5mm 20mm 160mm, 0mm}]{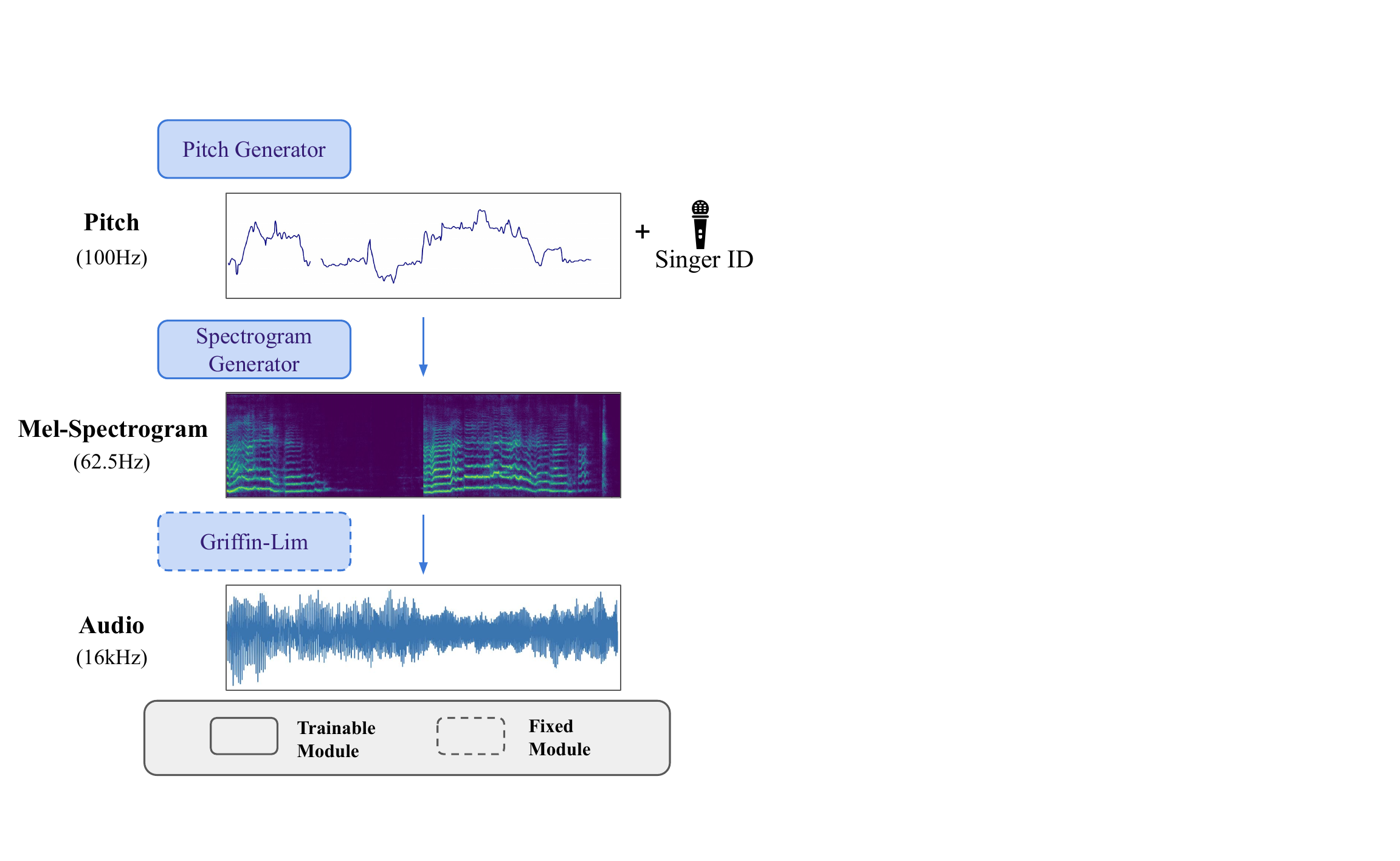}
    \caption{The hierarchical structure of the generative model, GaMaDHaNi comprising of a Pitch Generator, the Spectrogram Generator and a vocoder.
    }
    \vspace{-5mm}
    \label{fig:gamadhani}
\end{wrapfigure}
GaMaDHaNi (\cite{gamadhani}) is a two-level hierarchical generative model trained to sing vocal contours from Hindustani classical music. The model involves a Pitch Generator capable of generating a finely quantized fundamental frequency contour thus capturing the intricate melodic movements present in the musical form. Conditioned on this contour and singer information, a Spectrogram Generator predicts a mel-spectrogram which is further converted to audio using the Griffin-Lim algorithm (\cite{griffin1984signal}). Fig \ref{fig:gamadhani} presents a high-level representation of the model architecture. Through a listening study and qualitative analysis, the authors of GaMaDHaNi motivate their choice of the hierarchy and the intermediate pitch representation in the model architecture.

For the user study, a diffusion-based implementation of the Pitch and Spectrogram Generators was used, with the model interaction implemented via Gradio (\cite{abid2019gradio}). Further, I will briefly discuss the implementation of the tasks performed in the study. There are two main ideas involved (1) primed generation, which is used in the `idea generation' and `call and response' tasks and (2) melodic reinterpretation, which is seen in the `melodic reinterpretation' task. All of these concepts take place in the Pitch Generator.

\subsection{Primed Generation}
This approach involves prompting the model with a `prime', i.e. a melodic sequence represented as a pitch contour, and tasking it with generating a continuation. Since the model is trained to generate pitch contours of a fixed length $T=12$ seconds, we treat the last $t_{prime}$ seconds, where $t_{prime} < T$, of the prime as the initial part of the generated sample. The model then infills the remaining duration, $T - t_{prime}$ seconds, to complete the sequence.

For task 1, i.e. Idea Generation, the user’s only means of interaction is to choose from generated outputs of the model. Thus the samples were truncated to 4 s to give the user a sense of granularity and some control in creating a longer musical idea with this task. For each iteration of the loop, the duration of the prime considered, $t_{prime} = 2$ seconds. In task 2, i.e. Call and Response, the entire duration of the generation, i.e. 12 seconds was considered, where the first 4 seconds was taken from the input `call' or melodic prime, i.e. $t_{prime}=4$ seconds.

\subsection{Melodic Reinterpretation task}
\label{subsec:melodic-reinterp}
Melodic reinterpretation in this study mirrors the approach used in SDEdit (\cite{meng2022sdedit}) for guided image generation. This method takes advantage of the ability to solve reverse diffusion from any intermediate time step $t \in [0, 1)$. Using a coarse input or `guide', the appropriate initialization is found, using which the reverse diffusion process is performed to obtain a realistic pitch contour that is also faithful to the guide. 

GaMaDHaNi uses Iterative $\alpha$-Deblending (IADB) (\cite{heitz_iterative_2023}), as the training objective for both of its hierarchical blocks. IADB defines a simplified diffusion process that is a linear interpolation between noise $x_0 \sim X_0 = \mathcal N(0, 1)$ and data $x_1 \sim X_1 = X_{data}$. Thus an $\alpha$-blended point with blending parameter $\alpha$ is defined as, 
\begin{align}
    x_\alpha = (1 - \alpha)x_0 + \alpha x_1. \label{eq:stochastic-blending}
\end{align}

Thus for the model, given a melodic guide in the form of a pitch contour $x^{(g)}$, we find the proper initialization at time step $t_0$
\begin{align}
    x^{(g)}_{\alpha_{t_0}} = (1-\alpha_{t_0})x_0 + \alpha_{t_0}x^{(g)}, 
\end{align}
such that we can iteratively perform reverse diffusion with
\begin{align}
    x_{\alpha_{t+1}} = x_{\alpha_t} + (\alpha_{t+1} - \alpha{_t})D_\theta(x_{\alpha_t}, \alpha_t),
\end{align}

where blending parameter $\alpha_t = \frac{t}{T}$, $t \in [t_0, T-1]$ is the time step in the diffusion process, $T$ is the total number of steps, $x_{\alpha_t}$ is the value of the input $x$ at time t, and $D_\theta$ is the learned Pitch Generator model that is trained to predict the difference between the expected value of the posteriors of the data and noise samples given $\alpha$ and $x_{\alpha}$. For the purpose of this study $t_0$ is fixed at 0.5.

\subsection{Considerations while interpreting results from this ``in the wild'' study}
It is important to note that the generative model was trained on data from Saraga (\cite{srinivasamurthy2021saraga}) and Hindustani Raga Recognition (\cite{gulati2016time}) datasets, comprising studio and live ensemble performance recordings led by professional vocalists in Hindustani music. Thus, the data that the model sees during interaction in our studies is likely to be out of distribution for reasons including: (1) the input was less performance-like in terms of style and complexity, (2) training data was extracted from an ensemble setting whereas interaction data was from a solo participant, (3) the absence of source separation during interaction may have removed artifacts which were present in the training data, and (4) one of the participants used a harmonium for interaction while the model was trained only on the voice. Despite these differences, we conduct this exploration to inform future work to gear this model more toward interactive use cases.





\section{Challenges Faced by Participants}
\label{sec:challenges}

During the interaction, the model’s performance sometimes failed to meet the participants' expectations. Based on their reactions and discussions, we identified two primary challenges: (1) the lack of restrictions and (2) inconsistency in the model output. Readers are encouraged to watch samples of interaction for each of the pre-defined tasks and the challenges discussed below in the supplementary website.

\subsection{Lack of restrictions}
\label{sec:lack-of-restrictions}
Throughout all tasks, the participants noted a lack of restrictions on the output of the model. P2 claims that the model output was too ``free-flowing'' for him to evaluate musically, adding that ``the music that we perform has a very defined structure to it. And if it (the model) is going to be a companion, it cannot do some basic errors which are not there in the structure''. 

Participants felt the need for restrictions in the model output based on (1) \textit{raga}, (2) scale, (3) timbre, and (4) style as presented below.

\textbf{Raga}\space\space \textit{Raga} is an integral part of Hindustani music which defines the boundaries within which melodic improvisation can occur. \cite{powers-india} describe \textit{raga} as a combination of a scale and a tune; while it has defined notes and tonal hierarchies, it also includes characteristic phrases and melodic movements. Students of Hindustani music spend years understanding how to explore melody creatively while remaining within the framework of a \textit{raga} (\cite{mcneil2017seed}). The participants' initial observation was the lack of \textit{raga} adherence in model outputs (``But it's like just doing anything random'' (P1), ``it's not generally following any \textit{raga} rules'' (P2)), and consequently, P2 disfavored the samples for his practice saying, ``it would be more useful (for me) if it follows \textit{raga} rules''. Additionally, P3 constantly tried to make sense of the \textit{raga} that the model’s output was presented in by guessing the \textit{raga} the output was closest to (“it (the model output) has a vague Lalit (a \textit{raga} name) sensibility to it”). In tasks involving user audio input, P3 would try to use \textit{raga} as a means to guide the model output to stay similar to previous generations or even to ``shake things up'' to get different-sounding responses. However, he notes that this method may not be as effective stating that ``there's a little bit of potluck involved'' in getting \textit{raga} consistent input and output.
    
\begin{figure}
    \centering
    \includegraphics[width=\linewidth, trim={0, 0, 0.5cm, 0}]{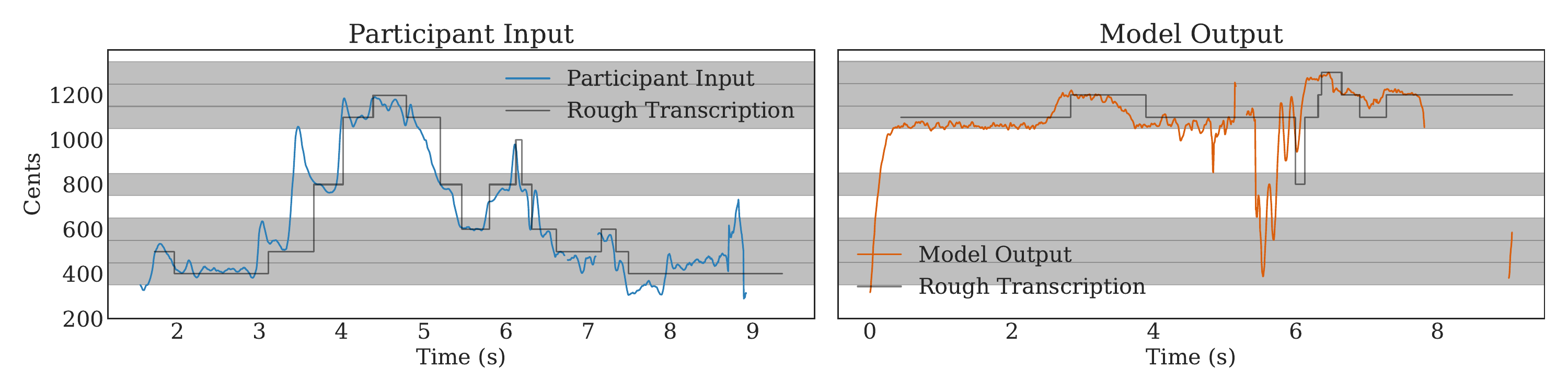}
    \includegraphics[width=\linewidth,trim={0.6cm, 8cm, 0.5cm, 8cm}]{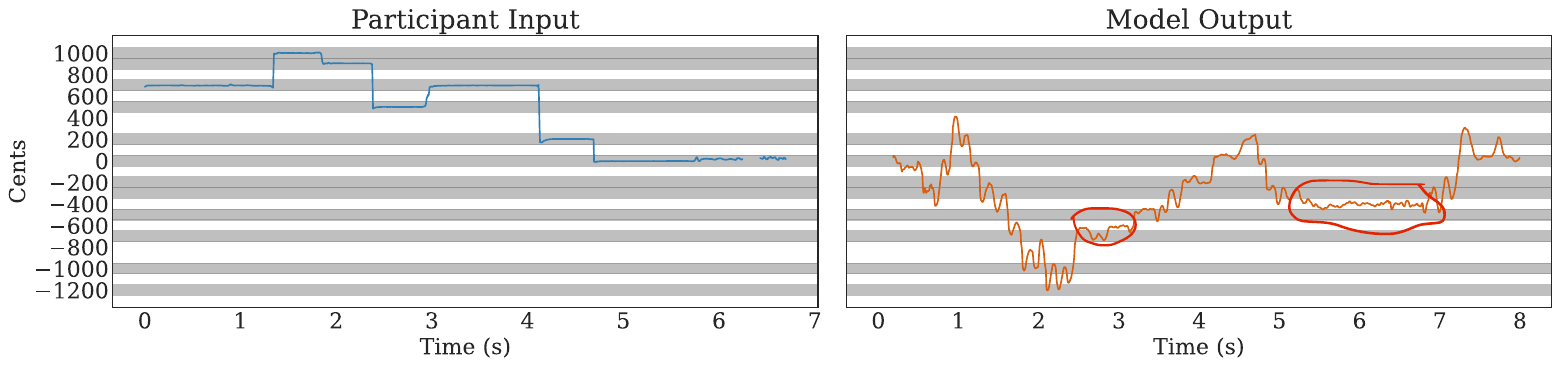}
    \caption{Extracted pitch contours of call and response examples where coherence was maintained (top) and not (bottom). \textbf{Top}: The model output maintained the scale that the input included. The scale notes used in the input are highlighted in horizontal grey boxes and a rough transcription performed by the author, a trained Hindustani musician, is provided as a black line to give a clear picture of the used notes. \textbf{Bottom}: An example of a model output that P1 found to have a `different structure'. One can see that while the input is very simple, the output has much more movement. Additionally the use of notes outside the scale (highlighted in red) results in incoherence as well. 
    }
    \label{fig:call-response}
\end{figure}

\textbf{Scale}\space\space While \textit{raga} is more than just a scale, P1 and P2 noted that even a simpler scale-based restriction would enhance their experience of interaction. For the call and response task, emphasizing the default expectation of scale-adherence, P2 notes ``it was not generally a good continuation, not even in the same scale''. On his preference of model outputs, P2 adds, ``... in one case I think it (the model) got 2 notes from the same scale (as the input). Yeah, it sounded much better than anything else.'' P1 views restrictions, specifically scale restrictions as a good thing taking the example of how one has to be careful when using a chromatic scale due to the 12 note options in an octave when compared to using simply a major or minor scale with only seven notes in an octave. Fig \ref{fig:call-response} highlights examples of where the scale was more or less maintained (top) and where it was not (bottom).

\textbf{Timbre}\space\space Due to the model's modular hierarchy, although the Spectrogram Generator of the model is conditioned on a singer's timbre, its effect holds good only if the pitch conditioning is in the range of the singer’s data. Thus, when the generated pitch contour is outside the singer's range, the timbre of the voice sounds different. All the participants noted this timbre shift and reacted differently. While P1 tried to modify his input to solicit the model output in the pitch range of the singer conditioning, P3 noted the delight in having a female-sounding (higher pitched) voice to interact with. P2 however pointed out that in performance, the timbre of a voice or instrument remains constant and thus should ideally remain constant during interaction as well.

\textbf{Style}\space\space Both P2 and P3 noted that the model’s output sometimes sounds like Carnatic\footnote{Carnatic music is a tradition of classical music from South India. While maintaining similar core concepts such as melodic and rhythmic frameworks, \textit{raga} and \textit{taal}, this music is aesthetically and stylistically different from Hindustani, which is a North-India classical music tradition.} music while expressing a desire to have model output that is more Hindustani-like as their input was consistent with that form. Given that the model was trained only on Hindustani music, future work will have to study what exactly makes generated samples sound more `Hindustani' or `Carnatic'.

\subsection{Incoherence in model output}

In the call and response task, participants often experienced a disconnect between their input and the model’s output, feeling that the model did not truly "respond" to their input, thus hindering two-way interaction. P1 notes that the outputs were ``random'' adding that his input did not seem to affect the output (`` … me giving input, I didn't see how it's helping''). Fig \ref{fig:call-response} illustrates instances where the model’s output seemed coherent or disconnected. Participants suggested that for the interaction to feel reciprocal, it should incorporate both lower-level technical attributes like \textit{raga} and scale, and higher-level abstract attributes like mood and musical idea.

\textbf{Low-level attributes}\space\space P1 notes how the model did not maintain the same musical ``structure'' as the input, taking the example of how he played one scale's notes but the model’s outputs were in another scale. Although the participant only gives the example of scale, I believe the idea of ``structure'' could also be extended to the rhythmic aspects of the input and output; the model often generates outputs in widely varied tempos from the input which can be disconcerting for a participant, an example of which is seen in Fig \ref{fig:call-response} (bottom). Additionally, P3 speaks of how he was singing the basic form of a \textit{raga}, Lalit, and thus expected the model to maintain the \textit{raga} but that was not the case.

\textbf{Higher-level attributes}\space\space P1 mentions how the model was uncooperative with his attempts to create a certain mood (``I was trying to achieve some mood and it gave me a different mood. So it didn't help because I would have to regenerate something (the input) to preserve the intended mood.''). Additionally, P2 remarks that the model did not give ``good responses'', which among other things discussed in Section \ref{sec:lack-of-restrictions}, hints at the model’s inability to capture the musical idea. Based on his practice, a ``good response'' involves the repetition of the musical idea in the input (“because the music that we do, it's a lot of repetitions. So usually the responses that I know of are repeating the same thing (as in the input).”, P2).

\section{Discussion}
Based on the observations from Sec. \ref{sec:challenges}, we note the importance of recognizing what creativity means in the context of Hindustani music and consequently, possible methods through which a model could better capture that essence.

\textbf{Creativity in Hindustani Music}\space\space \cite{nikrang2024ai} argue that creative systems should be able to produce surprising or unpredictable output autonomously while non-creative systems or tools are predictable and not necessarily autonomous. Hindustani music however involves idiomatic improvisation; the improvisation involved is not simply variations of a fixed composition itself, but elaborations of different \textit{raga} aspects performed against the \textit{taal}, i.e. metre (\cite{jairazbhoy1995rags}). Additionally, previous work (\cite{huron2008sweet, widdess2013schemas}) speaks of unconscious expectations among listeners in Hindustani music including schematic expectations, which are based on conventional patterns or structures in the music observed by the listener through long-term exposure to the music, and dynamic expectations which are based on the perception of patterns in real-time during performances. Thus one could argue that some predictability is required to satisfy conventional structures and frameworks. Further, \cite{mcneil2017seed} defines creativity in Hindustani music as the ability to expand and grow fixed seed ideas such as \textit{raga}, performance frameworks, and melodic ideas. Finding new melodic movements while satisfying these essential fixed seed ideas is necessary to create within the tradition.

\textbf{Constraints for creativity}\space\space Work by \cite{haught2017green} suggests that creativity thrives under constraints. Their study found that when participants were required to include concrete nouns while writing rhymes, the results were more creative. Constraints reduce the overwhelming array of possibilities to a more manageable subset, fostering imaginative thinking. Similarly, as P1 suggested, limiting the notes available to the model for generation could also encourage more interesting explorations. 

RL Tuner (\cite{jaques2017tuning}) employs reinforcement learning (RL) to enhance the output of a music sequence predictor, Note-RNN (\cite{eck2002finding}), using music theory-based constraints. Trained on MIDI data, which includes discrete note information, RL Tuner shows RL's potential for imposing restrictions in music generation. However, GaMaDHaNi, trained on continuous pitch contours, faces challenges in extracting discrete notes due to the subjectivity of solfege notation and the presence of note ornamentations and microtonal variations (\cite{vidwans2020classifying}). Thus, to effectively apply constraints in the discrete note domain, a robust mapping between discrete notes and continuous pitch contours must first be established.

AI researchers generally think of ideas of these constraints in terms of “conditioning”. Conditioning allows the user to shape the model output in desired ways, thus implicitly imposing the user's preferences. Conditioning could be performed either before, during, or after training a generative model. \textbf{First}, controlling the type of input that the model sees could ensure generations of only a certain data distribution. COCONET (\cite{huang2019counterpoint}), for instance, is trained only on Bach's chorales and thus generates music that harmonizes similarly. \textbf{Second}, conditioning could be added during training as an additional input. This could either be time-agnostic resulting in global conditioning or time-specific resulting in local conditioning. Jukebox (\cite{dhariwal2020jukebox}) involves global characteristics such as genre, artist, and style to condition model output whereas C-RAVE (\cite{devis2023continuous}) involves conditioning on time-varying local attributes such as loudness (measured with root mean squared energy), spectral brightness (measured through spectral centroid), sharpness and boominess (provided by AudioCommons project (\cite{audio-commons})). A common method of conditioning includes Classifier-Free Guidance (\cite{ho2022classifierfree}) which involves combining the outputs of the conditionally and unconditionally trained models to regulate the level of control. \textbf{Third}, post-hoc methods are used to instill conditions in pre-trained unconditional models, through learning constraints on the latent space of a model (\cite{engel2017latent}), prompt-based learning (\cite{brown2020language}), or training another model to impose constraints (\cite{zhang2023adding, wu2024music}). 

Thus, future work has to address effective ways to represent musical expectations in Hindustani performance such as \textit{raga}, \textit{taal}, and performance frameworks to aid creativity guided by the tradition. Further, studies will have to be conducted on methods to condition generated output on these expectations. Additionally as observed by the participants, further study is required to ensure the coherence of model output to the input to ensure a better interactive experience; this could potentially be implemented through various forms of conditioning, and accounting for the difference in data distributions seen during training and interaction.

\bibliographystyle{abbrvnat}
\bibliography{help}
\end{document}